# Urchin-inspired zinc oxide as building blocks for nanostructured solar cells

Jamil Elias,*† Mikhael Bechelany,‡ Ivo Utke†, Rolf Erni,†† Davood Hosseini,††† Johann Michler,† and Laetitia Philippe†

†EMPA, Swiss Federal Laboratories for Materials Science and Technology, Laboratory for Mechanics of Materials and Nanostructures, Feuerwerkerstrasse 39, CH-3602 Thun (Switzerland)

†† EMPA, Swiss Federal Laboratories for Materials Science and Technology, Electron Microscopy Center, Ueberlandstr. 129, CHD-8600 Dübendorf (Switzerland)

††† EMPA, Swiss Federal Laboratories for Materials Science and Technology, Laboratory for Thin Films and Photovoltaics, Ueberlandstr. 129, CHD-8600 Dübendorf (Switzerland)

‡ Institut Européen des Membranes (UMR CNRS 5635), Université Montpellier 2, Place Eugène Bataillon, 34095 MONTPELLIER, France

Corresponding Author

* Laboratory for Mechanics of Materials and Nanostructures EMPA Materials Science & Technology, Feuerwerkstrasse 39, 3602 Thun, Switzerland

Tel: +41 58 765 62 17. Fax: +41 33 228 44 90. E-mail: jamil.elias@empa.ch

ABSTRACT. High surface area nanowire based architectures have been identified as important components for future optoelectronic nanodevices, solar cells, wettability coatings, gas sensors, and biofuel cells. Here we report on a novel urchin-inspired nanowire architecture: its interwoven three-dimensional, high-surface-area nanowire arrangement can be precisely controlled by using a low-cost and scalable synthesis based on a combination of nanosphere lithography, low-temperature atomic layer deposition, and electrodeposition. The performance of single-layer arrays of urchin-inspired ZnO nanowire building blocks competes to that of planar nanowire carpets. We illustrate this capability by fabricating fully-inorganic extremely thin absorber solar cells using CdSe as absorber and CuSCN as hole-collector material. The light diffusion of the urchin-inspired nanowire arrays was varied from 15% to 35%. Homogenous absorption in the wavelength range of 400-800 nm of up to 90% was obtained. Solar conversion efficiencies of ~ 1.33% were achieved.

According to recent studies on the global power plant market, the installed capacity of solar power grew faster than that of any other power technology. Last generation nanostructured photovoltaic devices include dye sensitized (photoelectrochemical, quasi-solid, and solid-state) solar-cells[1] and their hybrid[2] and fully inorganic variants[3] as extremely thin absorber (ETA) solar-cells. They appear to have a big light harvesting potential compared to planar thin film photovoltaic devices due to their "built-in" large surface area architecture involving an n-type semiconductor material covered by a light absorber (dye, organic or inorganic films) for collecting photons. After charge separation, electrons are collected by a photoanode for electricity generation.

$TiO_2$ and ZnO were agreed to be the most promising materials as wide band gap n-type semiconductors with a preference for ZnO due to its better electronic transport properties and its comparatively easy controllable growth as single-crystal nanowire arrays[4-7]. Better control of light-scattering and electronic transport through this n-type semiconductor is essential for improving the solar efficiency[8]. Among numerous studied architectures, nanoparticles[1, 9] and nanowires[10] are the most employed building-blocks because they either provide high surface areas (nanoparticles) or direct electron transport (nanowires). In direct comparison, single-crystal nanowire arrays offer shorter electron collection paths, thus avoiding charge recombination[11]; but solar cells based on nanoparticles still have a higher solar efficiency[12] due to their larger surface area. Hence, increasing the surface area of planar nanowire carpets by increasing the diameter and length of the individual nanowire has been proposed in many research reports to enhance the solar light harvesting[10, 13]. As a commonly acquired result, such an increase of the surface area in nanowire carpets leads to an augmentation of charge recombination[10, 13] being detrimental for solar cell efficiency. Therefore, future nanostructured solar-cell architectures need to improve multiple light-scattering while keeping reasonable surface areas with a short electron collection path; in other words, improving the solar light absorption and reducing the electron-hole recombination. To tackle this challenge we have recently developed urchin-like nanostructures by electrodeposition of ZnO nanowires onto surface activated polymer spheres[14]. This structure showed a twofold improvement of light scattering compared to nanowire arrays. However, these nanostructures had a limited mechanical stability and their interspacing could not be varied which prohibited further optimized use in applications. In the present paper, we report on a novel architecture – based on a self-stabilized hollow urchin-like ZnO nanowire building-blocks using a novel low-cost and scalable synthesis route which allows for controlled building-block interspace and tunable nanowire dimensions. We show that the light diffusion and absorption as well as solar cell efficiency can be elegantly controlled and enhanced by engineering the dimensions of such building-blocks.

Controlled spacing and dimensions of hollow urchin-like ZnO nanowire (u-ZnO) building-blocks were obtained by combining electrodeposition together with polystyrene sphere (PS) lithography and atomic layer deposition (ALD). The schematic view in figure 1 summarizes the processes involved in the formation of u-ZnO (Figure 1I) as well as the fabrication steps for the final ETA solar-cells (Figure 1II).

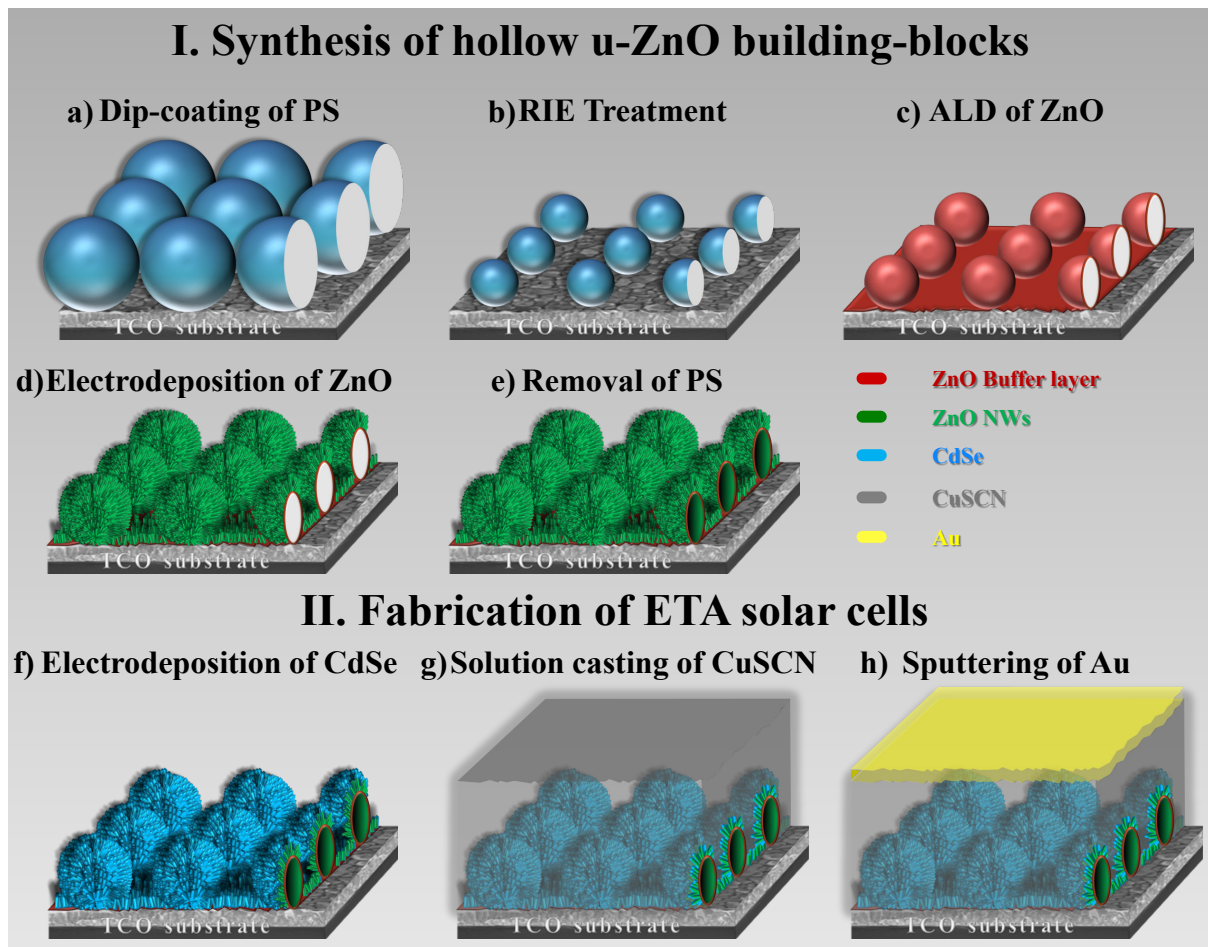

**Figure 1.** Schematic view of synthesis route for (I) self-stabilized hollow urchin-like ZnO nanowire building blocks and (II) successive fabrication steps for the ETA solar-cell: a) Dip coating for the deposition of an ordered monolayer of polystyrene microspheres onto an FTO covered glass substrate; b) Size reduction of spheres using reactive ion etching with oxygen plasma.; c) Deposition of a uniform conformal thin layer of about 20 nm of ZnO by ALD; d) Electrodeposition of n-type ZnO NWs with controlled length and diameter; e) Formation of hollow u-ZnO by by dissolving the polystyrene spheres in toluene; f) Coating of NWs with an absorber film of CdSe by electrodeposition; g) Covering with p-type CuSCN by chemical impregnation, and h) Deposition of a gold thin film electrode by physical vapor deposition.

After dip-coating of an ordered monolayer of PS onto a fluorine doped tin oxide (FTO) glass substrate (Figure 1a), the size of the spheres was reduced by a reactive ion etching (RIE) treatment which led to a homogeneous reduction of the spheres by maintaining their initial shape (Figure 1b). The size of the spheres was very accurately controlled by choosing the appropriate treatment time. Contrarily to the traditional use of the polystyrene lithography as nonconductive masks for deposition[15, 16], here the template will be used as a three-dimensional electrode by employing low-temperature ALD to obtain a uniform conformal thin layer (~20 nm) of ZnO on the three-dimensional ensemble of the FTO substrate and PS (Figure 1c). Indeed, this thin ALD ZnO layer plays three important roles: a) it adheres PS to the FTO substrate which needs to be immersed in a further step in an aqueous solution for electrodeposition, b) it renders the surface electrically active for the use as a conductive electrode for electrodeposition, and c) it is a barrier layer for blocking holes avoiding short-circuits between the FTO electrode and the p-type material[17]. In the next step, the FTO/PS/ZnO$_{ALD}$ ensemble was used as cathode for the electrodeposition of ZnO NWs (Figure 1d) employing the molecular oxygen electroreduction method[18, 19]. Consequently, hollow structures were obtained by removing PS in toluene or by oxidation in air at 450 °C (Figure 1e). We found out that this step can also be done directly after the ALD process. The remaining steps (Figure 1II) comprise the full process to obtain a solar-cell and will be referred to later in this article. In the following we will turn to the control of the specific geometry of these novel u-ZnO building-blocks and the effect of this architecture on light diffusion.

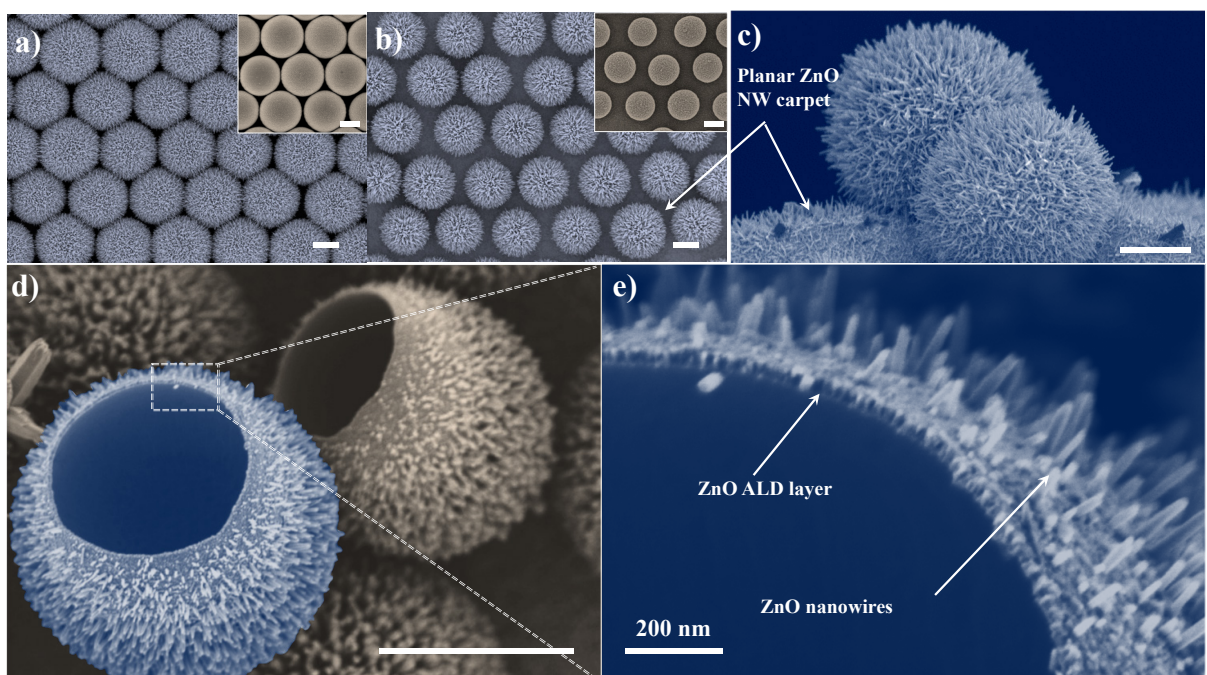

**Figure 2.** SEM images of ZnO urchin-like structures after dissolution of the PS monolayers a) without RIE and b) with 20 min RIE treatment. The insets of a) and b) are the SEM images of the ordered PS before electrodepostion coated with 20 nm of ZnO by ALD. c) Side view of individual u-ZnO structures. Note: the planar NW-carpet between the u-ZnO. d and e) are views of individual hollow u-ZnO structures from a scratched part of the sample where the structures were reversed upside down. All the scale bars in the figure (except (e)) are 2 μm.

Figures 2a and 2b show the ordered urchin-like structures obtained after electrodeposition of ZnO nanowires and the dissolution of PS. The nanowire arrays grew homogeneously on the ALD ZnO film covering the whole surface of PS leading to the formation of u-ZnO building-blocks. The insets of these figures show the ordered PS covered with ZnO ALD layers without and with RIE treatment (Figures 2a and 2b, respectively). The size of the spheres decreased from ~4.3 to ~3.3 μm leaving ~1 μm spacing between spheres after 20 minutes of RIE treatment. SEM images of a wider range of interspace variation can be found in the Figure S1. From a mechanically scratched part of the sample (Figure 2c) we can see clearly that a nanowire carpet was formed between the u-ZnO building-blocks increasing the surface area of the planar FTO substrate. An inside view of an individual building-block shown in figure 2d reveals that the u-ZnO structures are completely hollow from inside. The homogeneity of the ALD layer with a thickness of ~20 nm and its strong mechanical stability after removing PS are visibly confirmed in Figure 2e.

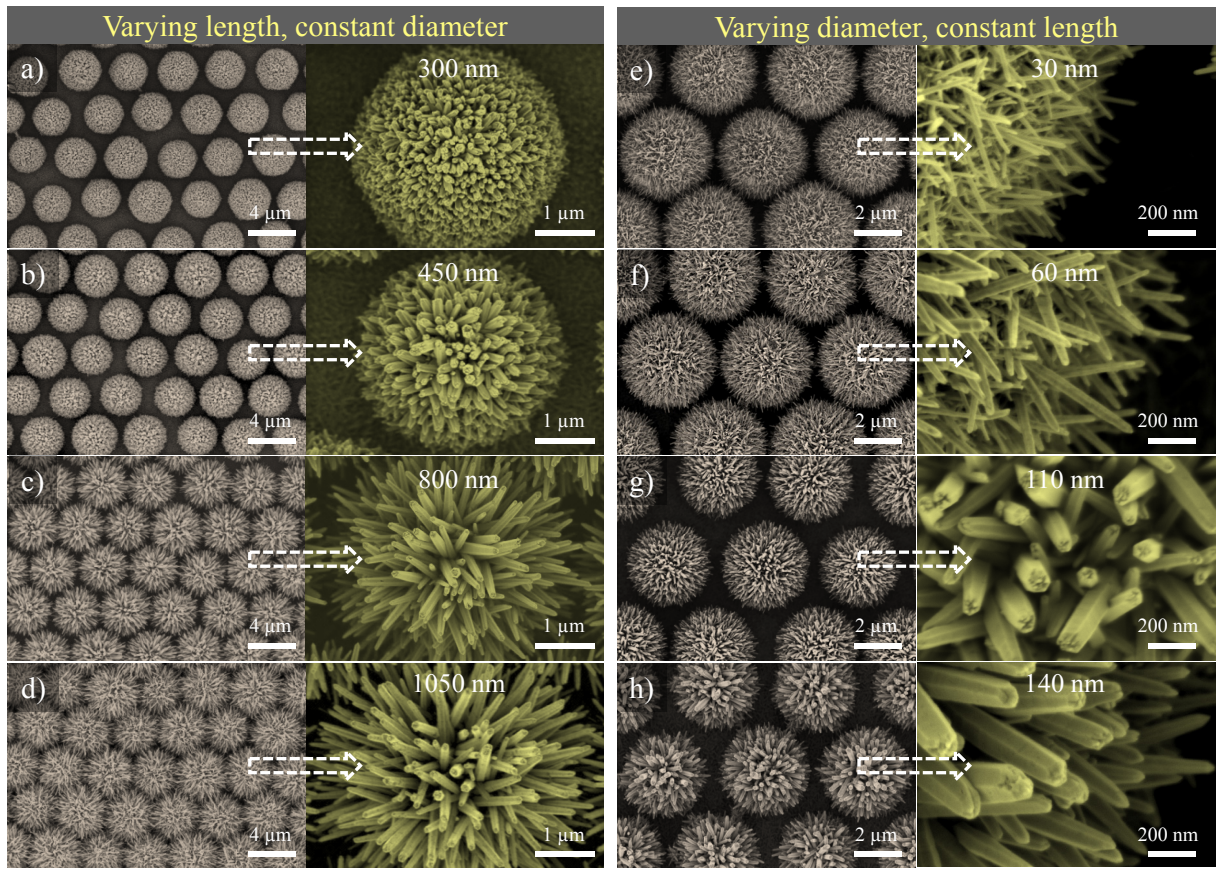

**Figure 3.** SEM images showing the dimension control of ZnO NWs composing the urchin-like building blocks. The RIE time and the ZnO ALD layer's thickness are fixed to 20 min and 20 nm, respectively. a-d) u-ZnO building blocks obtained at $5 \times 10^{-4}$ M $ZnCl_2$ and 0.1 M KCl with increasing electrodeposition charge densities from 5 to 40 $C.cm^{-2}$. e-h) u-ZnO building blocks obtained by increasing the concentration of $ZnCl_2$ from $5.10^{-5}$ to $1.10^{-3}$ M and keeping the KCl concentration at 0.1 M.

While the spacing and the overall size of the urchin structures can be controlled by the RIE treatment, the nanowire dimensions can be tailored by tuning the chemistry of the solution[20, 21]. Previously, we have shown that the electrodeposition from molecular oxygen is an efficient way to hinder the lateral growth of NWs and produce nanostructures with high aspect ratio[22]. As can be seen from the SEM images of figures 3, the longitudinal (Figures 3a-d) and the lateral (Figures 3e-h) growth of NWs were very precisely and *independently* controlled by tuning the electrochemical conditions. For both cases, the RIE time and the ALD layer's thickness were fixed to 20 min and 20 nm, respectively. In other words, the NW diameters were kept constant while varying their length in Figures 3a-d and vice-versa for the second series shown in Figures 3e-h. The high magnification SEM images confirm the regular variation of these dimensions. This study does hence prove the validity of the electrochemical growth mechanism of ZnO NWs on a three-dimensional substrate architecture.

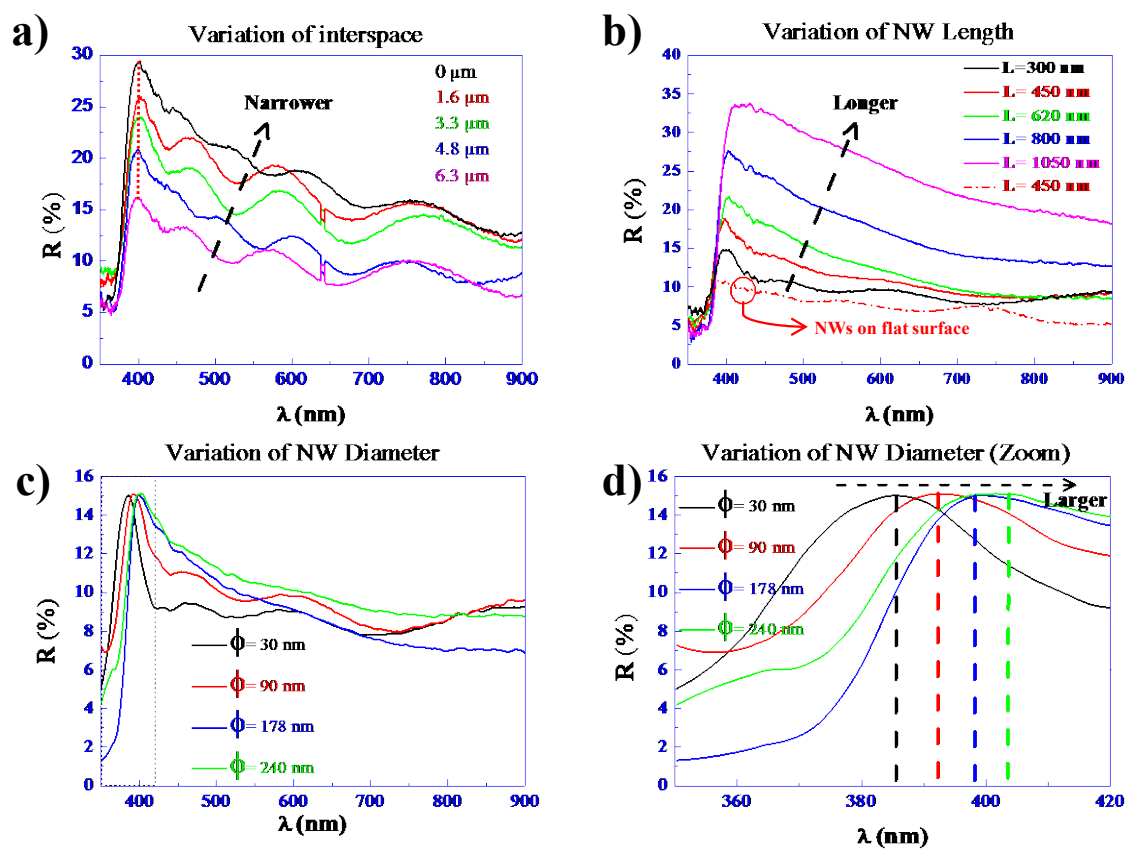

**Figure 4.** Optical reflectance of the samples obtained in figures 2 and 3: a) Variation of interspacing between the urchin-inspired building blocks; b) Increasing NW length by keeping constant the sphere interspace (3.3 μm) and the ZnO NW diameters (~100 nm); c) Varying NW diameters and keeping constant the NW lengths (~300 nm) and the sphere interspace (3.3 μm); d) Zoom into the region of UV absorbance edges of Figure 4c.

This novel approach allowed us to synthesize a wide range and numerous designs of ordered arrays of u-ZnO building blocks uniformly on relatively large surfaces (centimeter scale). As for planar ZnO carpet architectures composed of NWs,[23] nanotubes,[24, 25] and tetrapods[26], we strongly believe that the present urchin-like architecture can be used in a wide range of applications, like short-wavelength lasers,[27] piezoelectric nanogenerators,[28] electroluminescence,[29] and field-emission devices[30] apart from solar cells. We will proceed in demonstrating the feasibility of using u-ZnO architectures as an efficient building-block for ETA solar cells. A precise control of the geometry of the n-type semiconductor and a good understanding about the influence on light scattering are essential to improve the solar efficiency.[10] In order to study the geometry influence of u-ZnO building-blocks on the light scattering, the spectral dependence of the optical reflectance of the samples obtained in figures 2 and 3 has been measured and summarized in figure 4. For all analyzed samples, the spectra showed UV absorbance edges close to 375 nm (~3.3 eV), which corresponds to the band-gap energy of bulk ZnO.[31] The first set of measurements concerns the variation of interspace between the urchin-like building-blocks, the length (450 nm) and diameter (110 nm) of their nanowires being approximately constant (Figure 4a). By increasing the interspace, the maximum of the total reflectance (at $\lambda$~400 nm) decreased from 28% to 17% for interspaces from 0 to 6.3 µm along with a decrease of the reflectance over the whole wavelength range whereas only a marginal spectral peak shift was detected (dotted line, figure 4a). A similar behavior with more pronounced variation has been observed with an increasing NW length while the sphere interspace (3.3 µm) and the ZnO NW diameters (~100 nm) were kept constant (Figure 4b). The reflectance increased from ~ 15% to 34% (at $\lambda$~400 nm) when the NW length increased from 0.3 to 1 µm. In order to reflect the importance of the three-dimensional architecture based on urchin-like building blocks, a ZnO NW carpet (Length~450 nm, diameter~100 nm) was electrodeposited onto a flat FTO substrate that had been covered by a 20 nm thin ZnO ALD layer. The reflectance of this planar carpet architecture was measured (dashed curve, figure 4b) and compared to the corresponding urchin architecture having the same NW dimensions. An increase of about 50% of the reflectance was observed throughout the entire visible range. For constant NW lengths (~300 nm) and sphere interspace (3.3 µm), the variation of the NW diameter had no effect on the maximum reflectance value (Figure 4c), however a redshift with increasing diameter was observed (dashed lines, Figure 4d). The specific wavelength periodicities and dependencies found in figure 4 are subject to a theoretical treatment of multiple light scattering in arbitrarily oriented nanowire carpets in the frame of Mie's theory.[32] Here we give a qualitative interpretation by considering the surface fractions exposed to incident light of the u-ZnO arrays and the planar nanowire carpet surrounding them at their base. The planar nanowire carpets consist of NWs aligned preferentially vertical to the transparent planar substrate while u-ZnO constitutes a transparent hollow sphere with nanowires isotropically oriented in all directions (compare with figures 2 and S1). In figure 4a it is seen that the increase of the interspace between the u-ZnO building blocks decreases progressively the reflectance of such arrays. This tendency can be explained by reduced multiple light-scattering in this transparent material. With increasing interspacing a larger fraction of the planar nanowire carpet surface is exposed to the incident light while the surface fraction of u-ZnO nanowires inclined to the incident light diminishes. This explanation is strongly supported in figure 4b where a planar nanowire array shows about 50% less reflectance with respect to an u-ZnO array having the same nanowire dimensions. Figure 4b shows that increasing the nanowire length and keeping the interspacing constant led to two- to threefold larger reflectance. This is in line with explanation above: increasing the nanowire length decreased the light-exposed fraction of the nanowire carpet while the fraction of surface with nanowires *inclined* to the incident light increases. Furthermore, a recent experimental study[7] concerning the effect of ZnO nanowire size on multiple light scattering showed that the length of vertical nanowires in a planar carpet architecture increases the reflectivity only for a certain wavelength range (400-600 nm). In our urchin-like array architecture, the increase of the reflectivity is found to cover a *large* wavelength range (400-800 nm). This is most probably due to the fact that nanowires from urchin-like building blocks are exposed to the incident light from all angles. Figures 4c and 4d show that an increase in the nanowire's diameter results in a redshift of the spectra. Interestingly, the spectra do not differ largely in the long-wavelength range which can again be attributed to the fact that surface fractions of the NW carpet and u-ZnO exposed to incident light do not change considerably when changing the nanowire diameter.

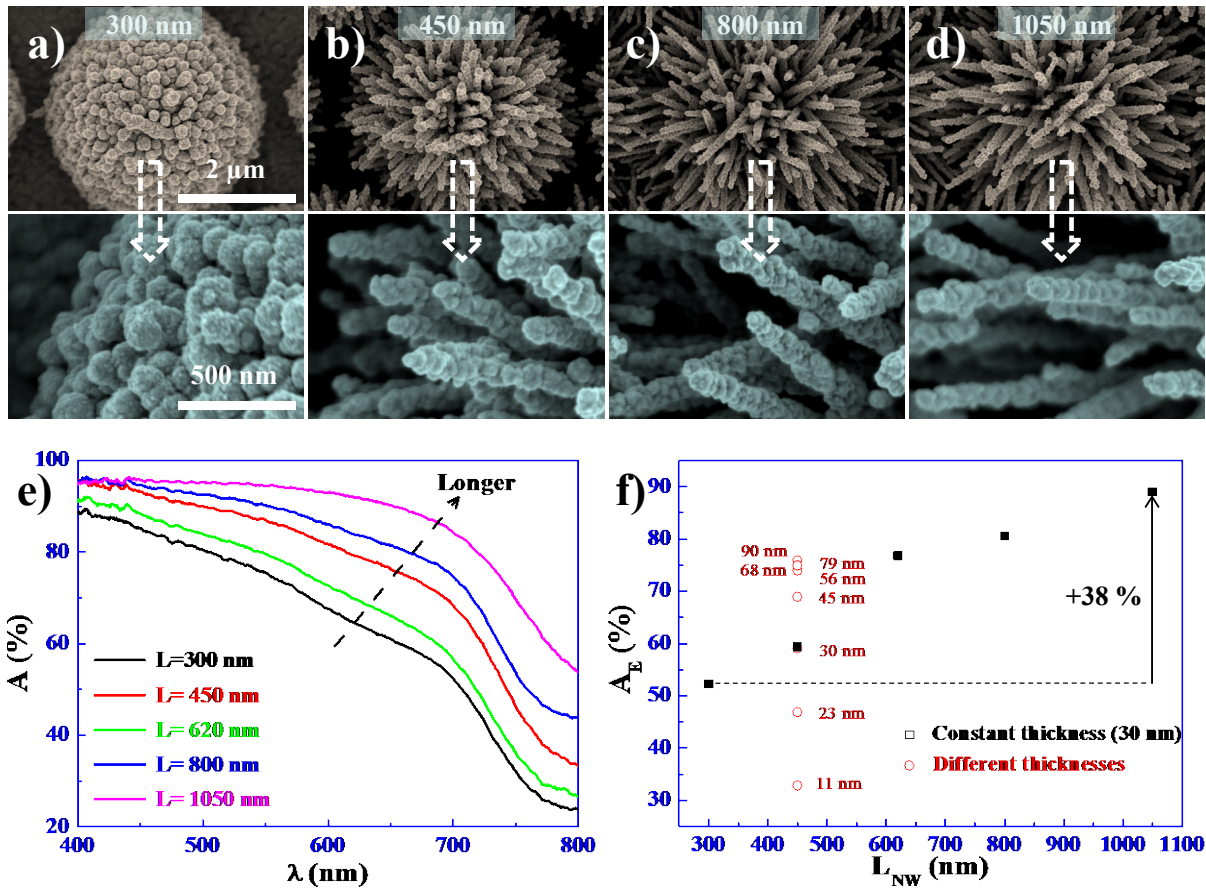

**Figure 5.** a-d) SEM images of core-shell ZnO/CdSe urchin-like building blocks with varying lengths having constant CdSe thickness and ZnO NW diameter; e) The absorbance (A) spectra of the CdSe coated urchin building blocks; and f) Effective absorption ($A_E$) of ZnO/CdSe urchin-like building blocks versus ZnO NW length having a constant CdSe thickness of 30 nm. Also shown is the effective absorption for a constant NW length (450 nm) and varying CdSe thicknesses.

To study the effect of light scattering on the sensitization of the urchin-like building-blocks we coated the series of u-ZnO building blocks with varying NW lengths (Figures 3a-d) with a CdSe absorber film of constant thickness of ~30 nm (Figure 1f). CdSe is a narrow band-gap (~1.7 eV) semiconductor and has been successfully used in different nanostructured solar-cells as a light absorber[3, 33]. Electrodeposited core-shell ZnO/CdSe structures with varying ZnO NW lengths (300-1050 nm) and a constant CdSe thickness (30 nm) are shown in Figures 5a-d. Since the ZnO NW diameters were constant (~100 nm), a uniform average ZnO/CdSe core-shell NW diameter of these structures is seen in the high magnification images of figure 5.

Figure 5e displays the absorbance (A) spectra resulting from measurements of the total transmittance (T) and reflectance of the CdSe coated urchin-like building-blocks (A=100-T-R). Compared to the reflectance graph (Figure 4b) a clear correlation was detected for the absorbance after adding the CdSe layer, indicating that multiple scattering in the urchin-like structures enhances the optical absorption through ZnO/CdSe core-shell. A quantification of these absorption spectra is made by calculating the effective absorption ($A_E$) in the region where CdSe absorbs (400-800 nm) and the results are displayed in the graph of Figure 5f. We determined the values of $A_E$ from the following equation[34]:

$$A_E = \int_{\lambda 1}^{\lambda 2} A(\lambda) J_0(\lambda) d\lambda \bigg/ \int_{\lambda 1}^{\lambda 2} J_0(\lambda) d\lambda$$

where $J_0$ is the photon flux. Figure 5f highlights the importance of the optical engineering of u-ZnO nanostructures on the light absorption. An increase of $A_E$ from ~52% to ~90% is already obtained by increasing the NW lengths from 0.3 to 1μm (enhancement of 38%). In order to confirm the NW dimension effect on the scattering and the absorption, we fixed the length of NWs to ~450 nm and varied the thickness of CdSe absorber layer from 11 to 90 nm. The red circles in Figure 5e indicate that even by increasing the thickness up to 90 nm, $A_E$ values were still lower than the one with the thinner CdSe layer (~30 nm) and longer NWs (~1 μm). This highlights that the absorption of ZnO/CdSe core-shell building-blocks can be efficiently enhanced by our approach.

The final ZnO/CdSe/CuSCN ETA solar-cells were obtained by covering the core-shell ZnO/CdSe urchin-like building-blocks with CuSCN and a gold sputtered layer (Figures 1g and 1h). CuSCN is a wide bandgap (~3.4 eV) p-type semiconductor which has been used previously to act as a hole collector in ETA solar-cells[3].

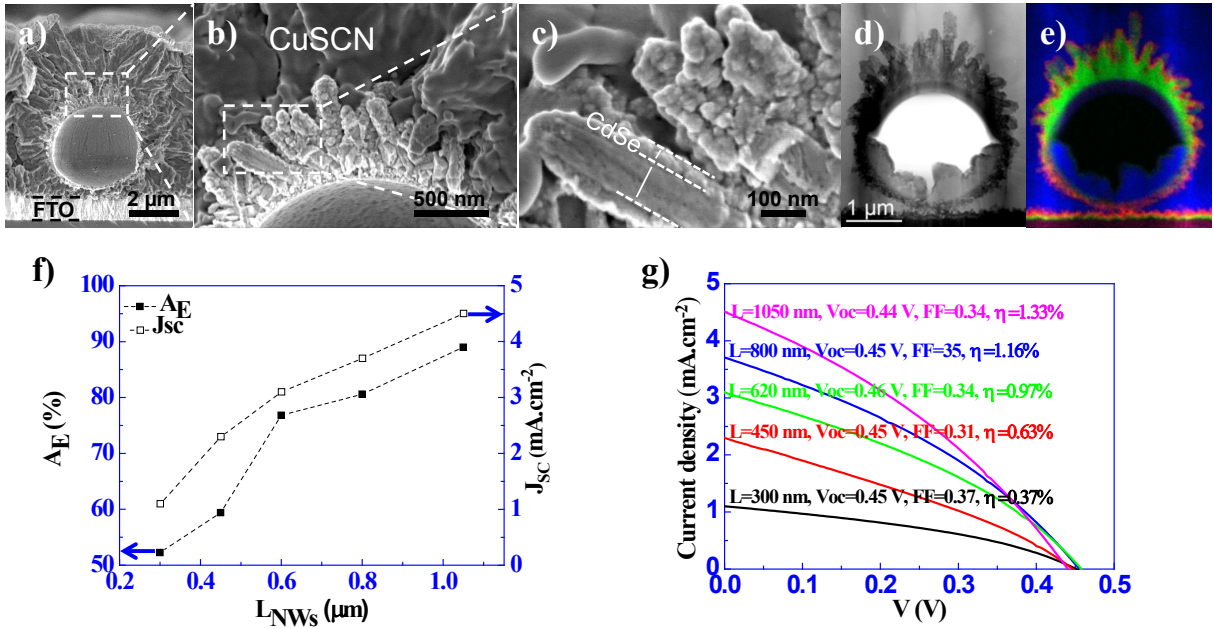

**Figure 6. a-c)** Cross-sectional SEM images of an individual urchin-like n-type ZnO building block (NWs: 100 nm diameter, 450 nm length) integrated in a fully processed solar-cell containing the CdSe absorber layer and p-type CuSCN; d) STEM image of the another building block (NWs: 100 nm diameter, length 1 μm) prepared as a 100 nm thin TEM lamella. The strong white contrast is due to the hollow nature of the building block.; e) EDX mapping made on (d), blue = Cu, purple = Cd, green = Zn, and black = vacuum. The presence of CuSCN inside the sphere is due to a preparation artefact; f) Short circuit current densities $J_{sc}$ of solar-cells based on urchin-like building blocks having different lengths of ZnO NWs. The solar cells are measured under half-sun (50 mW.cm$^{-2}$). For a direct comparison effective absorbance values ($A_E$) obtained in Figure 5 are displayed together; and g) Current-density curves including the characteristics of solar-cells cells based on urchin-like building blocks having different lengths of ZnO NWs.

Cross-sectional SEM images of a u-ZnO building-block in a fully processed solar-cell (diameter 100 nm, $L_{NWs}$~450 nm) are shown in Figure 6. Figure 6a shows a high impregnation quality of CuSCN from the bottom to the top of the urchin nanostructure. The zooms taken at the interfaces of the ZnO/CdSe/CuSCN heterostructure (Figures 6b and 6c) confirm the good diffusion of the CuSCN solution and crystallization within the narrow interspace of NWs. A mechanically scratched part of the core-shell ZnO/CdSe NW (dashed lines in Figure 6c) shows the high homogeneity of CdSe thickness (~30 nm) and the diameter of ZnO NW (~100 nm). Using a TEM cross-section lamella of another building block ($L_{NW}$~1 μm, figure 6d) for energy-dispersive X-ray mapping in a scanning transmission electron microscopy confirms that CdSe is homogeneously covering the urchin-like ZnO building-block as well as the NW carpet deposited directly on the FTO substrate. The same is true for the homogeneous infiltration by CuSCN.

Solar-cells based on urchin-like building-blocks with differing lengths of ZnO NWs were measured under 50 mW.cm$^{-2}$ illumination conditions and the characteristics are displayed in Figures 6f-g. It is clearly seen that the short circuit current ($J_{sc}$) values followed the same behaviour as the effective absorbance ($A_E$). The values of the open-circuit voltage ($V_{oc}$) and fill factor (FF) are almost constant with increasing NW length (see supplementary information). The combination of these characteristics leads to an almost linear increase of the conversion efficiency (Figure 6g) from 0.37% to higher than 1.33% by only increasing the ZnO nanowire lengths from 0.3 to 1 μm. These results prove, for the first time, that solar-cells can be made by three-dimensional, urchin-like ZnO nanowire building blocks. Although our actual best solar-cell efficiency is still slightly lower by (1.33% versus 2%) with respect to the few published works concerning ZnO/CdSe/CuSCN solar-cells[3, 12], we strongly believe that there is a big potential for solar cell improvement by further optimizing the geometry of urchin-like ZnO nanowire building-blocks as well as for the optimization of absorber layer and hole collector layer thicknesses and materials.

In conclusion, we have established a novel low-cost synthesis route to produce large areas of perfectly-ordered hollow self-stabilized urchin-like ZnO nanowire building-blocks by a combination of PS patterning, ALD, and electrodeposition. The process allows a wide control of the dimensions of arrays of urchin-inspired building-blocks and their constituting ZnO nanowires. It opens the door for investigations of a large class of semiconducting and metallic materials in more complex three-dimensional architectures for different type of applications, like short-wavelength lasers,[27] piezoelectric nanogenerators[28], electroluminescence,[29] and field-emission devices[30]. In this study we proved the functionality of this unique architecture in extremely thin absorber solar-cells for the first time. A clear correlation between the geometry of the urchin-like building block, the optical properties, and the solar efficiency has been stated offering promising applications of such building-blocks in different types of nanostructured solar-cells (organic, hybrid and dye sensitized solar-cells) as well as in applications where specific architectures of high surface area structures are an asset like for wettability, gas sensing, and biofuel cells. At this moment we are investigating applications of these urchin building blocks in novel enzymatic biofuel cells and nanowire-assisted laser desorption/ionization where very promising preliminary results were obtained.

# Methods

A commercially available polystyrene microsphere suspension (diameter ~ 4.3 μm, 4 wt. % aqueous dispersion) from Duke Scientific, U.S.A, was used for dip coating. Fluorine doped tin oxide (FTO) transparent conductive substrates ($SnO_2$:F, sheet resistance:10 Ω/sq, size 2x5 cm$^2$) were purchased from Xop Fisica S.L., Spain. The FTO glass substrates were consecutively cleaned thoroughly by sonication in acetone, ethanol, and isopropanol for 15 minutes each. An amount of 300 μL of polystyrene microsphere suspension was diluted with an equal volume of ethanol and dispersed (drop-by-drop using a micro-pipet) onto the surface of deionized water filled into a Petri dish of 8 cm diameter. After the self-assembling process of polystyrene spheres on the water surface, the FTO substrate was immersed into the solution. The polystyrene sphere monolayer was transferred onto the FTO substrate surface by slow removal from the solution at an angle of 45°.[14] After complete drying of the polystyrene sphere covered FTO/glass substrate, the regularly arranged, closely packed spheres were reduced in diameter by reactive ion etching (RIE) using an oxygen plasma treatment. The size of the polystyrene spheres was very accurately controlled by choosing the appropriate RIE time.

Consecutively, the polystyrene sphere covered sample was introduced in a custom made atomic layer deposition (ALD) reactor for synthesizing a thin ZnO film covering homogeneously the spheres and the $SnO_2$:F substrate surface. The ZnO ALD films were deposited using alternating exposures of diethylzinc (DEZ) and water ($H_2O$). The deposition was carried out at 60°C with the following cycle times: 0.1 s pulse (DEZ), 20 s exposure, and 45 s purge. A 1 s pulse, 20 s exposure and 60 s purge was used for $H_2O$. The time for pulse, exposure, and purge cycles were chosen conservatively to ensure saturation of the ALD surface reactions and to prevent inadvertent mixing of the reactive gases. To obtain a ~20 nm ZnO film 100 cycles were performed.

The sample, conformally covered with an ALD ZnO thin film, was then used as the working electrode in a three-electrode electrochemical cell. A platinum spiral wire was used as the counter electrode and a saturated calomel electrode (SCE) as the reference electrode. The electrolyte was an aqueous solution saturated by molecular $O_2$ containing zinc chloride as zinc precursor ($5x10^{-5}$-$1x10^{-3}$ M) and potassium chloride (0.1 M) as supporting electrolyte. The electrodeposition was performed at a constant electric potential (-1 V) and the charge density was varied between 5 to 40 $C/cm^2$ using an Autolab PGSTAT-30 potentiostat. The bath temperature was set to 80 °C. After electrodeposition, the polystyrene microspheres were dissolved in toluene or oxidized in air at 450 °C for 1 h. The above process steps defined the hollow urchin-like ZnO NW building blocks (hollow u-ZnO). For their use in solar cells, the following processes were employed:

The extremely thin absorber film of CdSe was electrodeposited onto the hollow u-ZnO NW building block following the method published in [35]. CdSe films were electrodeposited from an aqueous selenosulfate solution. The electrodeposition solution contained cadmium acetate 0.5 M ($C_4H_6CdO_4$, Sigma Aldrich, purity ~98%), sodium nitrilotriacetate 1 M (NTA, Fluka, purity > 98%), sodium sulfite 0.4 M ($Na_2SO_3$ Fluka, purity > 98%) and Selenium 0.2

M (Se, Riedel de Haën ~99%). Se was dissolved in 0.4 M $Na_2SO_3$ aqueous solution at 60 °C and stirred for 5h. After preparation, the solution can be used for few weeks. The pH was adjusted to 8.5 by diluted acetic acid solution (~ 3 %). The electrodeposition was carried out galvanostatically at room temperature in a two-electrode electrochemical cell. The working electrode was ZnO/FTO cathode and the counter electrode was a Pt spiral wire. The applied current density was -3 $mA.cm^{-2}$. In order to obtain a constant thickness of CdSe, different charge densities values have been estimated (0.2 - 0.8 $C.cm^{-2}$) depending on the estimated surface area of the u-ZnO building blocks having different NW lengths.

The hole-collector film was made of CuSCN and was synthesized by a chemical impregnation method.[36]

The metallic contact consists of a thin gold film of 100 nm thickness sputtered on top of the CuSCN. The above three final process steps defined the solar-cell investigated in this study. The morphology of the samples was observed by scanning electron microscopy (SEM, Hitachi S-4800). Transmission electron microscopy (TEM), scanning transmission electron microscopy (STEM), and energy dispersive X-ray spectroscopy (EDX) were performed on a JEOL JEM 2200fs microscope, operated at 200 kV. The TEM cross-section sample was prepared by mechanical polishing followed by argon ion milling. The reflectivity of the hollow u-ZnO NW building blocks and the absorbance of CdSe covered u-ZnO NW building blocks were characterized by a spectrophotometer from Ocean Optics (HR2000+ES) fitted with two integrating spheres (for transmittance and reflectivity). Current-voltage (J-V) measurements were made on the ETA-solar cells under half sun illumination (50 $mW.cm^{-2}$). The characteristic parameters of the solar cells, including short-circuit current density (Jsc), open-circuit voltage (Voc), fill factor (FF) and overall photoconversion efficiency were extracted from the J-V curves.

# Acknowledgment.

The authors thank Amaury Bruneau and Jeanne Baudot for valuable discussions and experimental help. This work was supported by SFOE (Swiss Federal Office for Energy) ETA Solar Cell project, No. 103296, Switzerland.

Supporting information

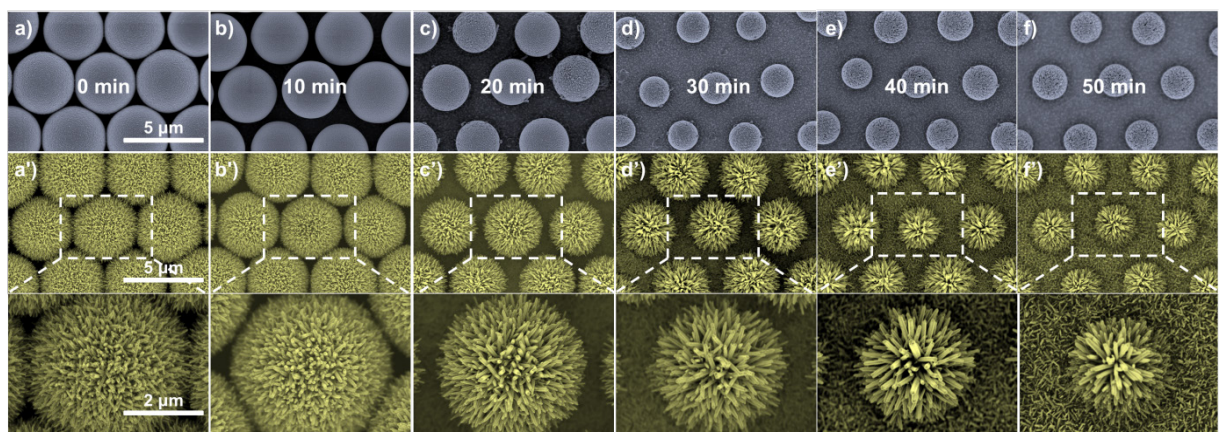

**Figure S1.** SEM images showing the effect of RIE time on the interspace between PS before after electrodeposition of ZnO nanowires. a-f) PS dimensions' variation as a function of RIE time. a'-f') urchin-like ZnO obtained after electrodeposition of ZnO ($5\times10^{-4}$ M $ZnCl_2$, 0.1 M KCl, and Q=10 $C.cm^{-2}$). The SEM pictures in the bottom are the high magnification of a'-f').